\documentclass[conference]{IEEEtran}

\usepackage{cite}
\usepackage{amsmath,amssymb,amsfonts}

\usepackage{graphicx}
\usepackage{booktabs}
\usepackage{multirow}
\usepackage{makecell}
\usepackage{tabularx}
\usepackage{enumitem}

\usepackage{listings}
\usepackage{tcolorbox}
\usepackage{hyperref}
\usepackage{tikz}
\usepackage{textcomp}
\usetikzlibrary{positioning,arrows.meta}

\usepackage{pifont}
\usepackage{textcomp}
\usepackage{xcolor}
\usepackage{verbatim}
\usepackage{url}

\newcommand{\cmark}{\ding{51}}

\newcommand{\spara}[1]{\smallskip\noindent{\bf #1}}

\makeatletter
\let\originalLstCaption\lst@makecaption
\renewcommand{\lst@makecaption}[2]{%
  \originalLstCaption{#1}{#2}%
  \vspace{5pt}%
}
\makeatother

\graphicspath{{figures/}}

\lstdefinestyle{schemastyle}{
  basicstyle=\ttfamily\scriptsize,
  columns=fullflexible,
  keepspaces=true,
  breaklines=true,
  showstringspaces=false,
  upquote=true,
  frame=single
}

\lstdefinestyle{promptstyle}{
  basicstyle=\ttfamily\scriptsize,
  columns=fullflexible,
  keepspaces=true,
  breaklines=true,
  showstringspaces=false,
  upquote=true,
  frame=single
}

\tcbset{
  vldbbox/.style={
    colback=white,
    colframe=black!60,
    colbacktitle=black!8,
    coltitle=black,
    boxrule=0.4pt,
    arc=0pt,
    left=2pt,
    right=2pt,
    top=2pt,
    bottom=2pt,
    fonttitle=\bfseries\footnotesize
  }
}

\begin{document}

\title{[Experiment, Analysis, and Benchmark]
How well do LLMs comprehend and design relational database schemata?}

\author{\IEEEauthorblockN{Sriom Chakrabarti}
\IEEEauthorblockA{
\textit{Aalborg University}\\
Denmark \\
scha@cs.aau.dk}
\and
\IEEEauthorblockN{Sean Bin Yang}
\IEEEauthorblockA{
\textit{Aalborg University}\\
Denmark \\
seany@cs.aau.dk}
\and
\IEEEauthorblockN{Arijit Khan}
\IEEEauthorblockA{
\textit{Bowling Green State University, USA}\\
\textit{Aalborg University, Denmark}\\
arijitk@bgsu.edu}
\and
\IEEEauthorblockN{Sebastian Link}
\IEEEauthorblockA{
\textit{University of Auckland}\\
New Zealand \\
s.link@auckland.ac.nz}
}
\maketitle
\pagestyle{plain}
\thispagestyle{plain}


\begin{abstract}
Large Language Models (LLMs) are increasingly used for database tasks, yet most work, such as \textsf{Text2SQL}, assumes that the relational schema already exists. We ask whether LLMs can design a schema from natural language requirements (\textsf{Text2Schema}), verbalize an existing schema (\textsf{Schema2Text}), and preserve its structure through a Schema$\rightarrow$Text$\rightarrow$Schema round trip. We present \textsf{SchemaBridge}, a benchmark of 902 schemata from three collections, stratified by size and constraint complexity, that scores fidelity separately for tables, attributes, domain (data types and nullability), entity, and referential integrity. Evaluating six open-weight and proprietary LLMs, we find that:
(i)~human-written requirements under-specify schemata, with no model exceeding 0.27 referential integrity F1;
(ii)~LLM-generated descriptions carry far more structure when explicitly asked to be complete, raising GPT-5.4-mini's referential F1 on \textsf{SchemaBench} from 0.228 to 0.909, even as BLEU, ROUGE-L, and BERTScore decrease;
(iii)~controlled experiments show that both constructor and describer choices influence round-trip fidelity under fixed settings; and
(iv)~masked schema experiments show that structural cues support recovery of keys and relationships, while joint data-type and nullability information is poorly retained through verbalization.
Code and data are available at \url{https://github.com/iamsriom/SchemaBridge} 
\end{abstract}


\section{Introduction}
\label{sec:introduction}

\begin{figure*}[t]
    \centering
    \includegraphics[width=0.94\textwidth]{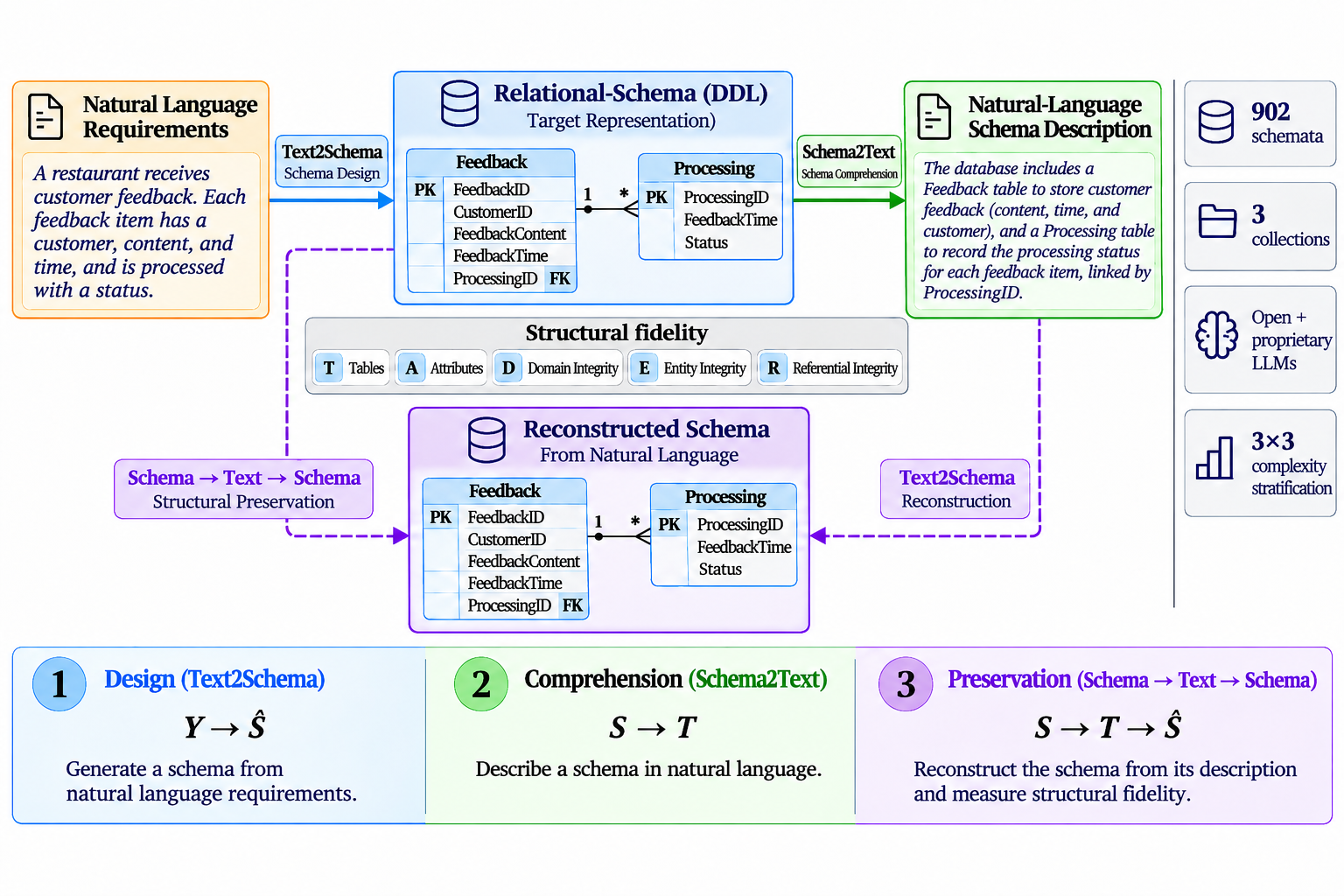}
    \caption{Overview of \textsf{SchemaBridge}. We evaluate LLMs across three complementary relational-schema reasoning tasks: schema design (\textsf{Text2Schema}), schema comprehension (\textsf{Schema2Text}), and structural preservation through the $\text{Schema}\rightarrow\text{Text}\rightarrow\text{Schema}$ round trip. Structural fidelity is measured over tables, attributes, domain integrity, entity integrity, and referential integrity.}
    \label{fig:schemabridge-glance}
\end{figure*}

Large Language Models (LLMs) are increasingly being used to interact with relational database systems, yet most evaluations assume that the schema is already provided. In \textsf{Text2SQL}~\cite{gao2024text,PourrezaL0CTKGS25,KoukouvinisKK26}, for example, a natural language question is translated into an executable SQL query over a provided relational schema, with benchmarks such as WikiSQL~\cite{zhong2017seq2sql}, Spider~\cite{yu2018spider}, and BIRD~\cite{li2023can} driving substantial progress in this setting. The complementary \textsf{SQL2Text} task
~\cite{xu2018sql,camara2024large,al2025semantic,chakrabarti2025sql} verbalizes individual SQL queries rather than complete relational schemata. These lines of research demonstrate that language models can operate over supplied relational structure but do not establish whether they can construct the schema itself or faithfully express the relational structure encoded by an existing design.

Natural language and relational schemata represent information at different levels of abstraction. Natural language descriptions usually express application concepts, business rules, and domain semantics, but many schema details such as tables, attributes, data types, domain constraints, and referential constraints may remain implicit or underspecified. In contrast, Data Definition Language (DDL) explicitly encodes this structural information but provides limited information about the business rationale or domain meaning behind it. This difference makes translation between the two representations challenging. When generating a schema, an LLM may need to infer structural details that are not explicitly stated in the requirements, while during schema comprehension it must preserve enough relational information in natural language for the original structure to remain understandable and recoverable.

These differences lead to several challenges when evaluating whether LLMs can faithfully reason over a relational schema. First, natural language requirements are often incomplete, so an LLM needs to infer schema elements that are not clearly stated. Second, the same requirements can admit multiple valid relational designs, which makes exact match with a reference schema a stricter criterion than functional necessity. Third, when an existing schema is expressed in natural language, important structural information such as constraints and relationships may be omitted even when the description appears fluent and complete. Finally, in a Schema$\rightarrow$Text$\rightarrow$Schema pipeline, structural errors may arise either during schema verbalization or during reconstruction, making it difficult to determine where information is lost. Existing \textsf{Text2SQL} and \textsf{SQL2Text} benchmarks do not evaluate these challenges at the level of the complete relational schema, while recent \textsf{Text2Schema} work \cite{wang2025schemaagent} primarily considers schema construction in one direction. To the best of our knowledge, no existing benchmark jointly evaluates schema construction, schema comprehension, and structural preservation across natural language and relational schema representation. It therefore remains unclear how well LLMs can construct, express, and preserve relational structure across natural language and formal schema representations. These challenges motivate an evaluation that measures not only the quality of generated schemata or descriptions but also which parts of relational structure are preserved, omitted, or incorrectly reconstructed across the two representations.

To address these challenges, we introduce \textsf{SchemaBridge}, a unified benchmark for evaluating relational schema reasoning across natural language and formal schema representations. As illustrated in Figure~\ref{fig:schemabridge-glance}, \textsf{SchemaBridge} considers three complementary evaluation settings: schema design (\textsf{Text2Schema}), schema comprehension (\textsf{Schema2Text}), and structural preservation through a Schema$\rightarrow$Text$\rightarrow$Schema round trip. The benchmark evaluates 902 relational schemata from three complementary collections using multiple open-weight and proprietary LLMs. Rather than relying on a single aggregate metric, \textsf{SchemaBridge} measures structural fidelity across five dimensions: tables, attributes, domain integrity, entity integrity, and referential integrity. This fine-grained evaluation enables us to identify where relational information is preserved or lost and to distinguish errors arising during schema construction, verbalization, and reconstruction.

We make the following contributions:

\begin{itemize}[leftmargin=*]
  \item \textbf{Unified benchmark.} We develop \textsf{SchemaBridge}, covering schema design (\textsf{Text2Schema}), schema comprehension (\textsf{Schema2Text}), and structural preservation (Schema$\rightarrow$ Text$\rightarrow$Schema) over 902 schemata from three collections, stratified by size and constraint complexity (\hyperref[sec:methodology]{\S\ref*{sec:methodology}}--\hyperref[sec:experimental_setup]{\ref*{sec:experimental_setup}}). 
  
  \item \textbf{Diagnostic evaluation.} We measure fidelity separately for
  tables, attributes, domain, entity, and referential integrity. We also perform controlled experiments 
  to isolate the effects of the constructor and the describer, while we perform masking to assess the recovery of hidden schema elements from visible structural evidence. (\S\ref{sec:methodology}).

  \item \textbf{Empirical findings.} Across six LLMs, we find:
    (i)~limited exact-match fidelity when constructing schemata from provided requirements;
    (ii)~improved round-trip preservation under detailed descriptions;
    (iii)~variation in reconstruction performance across describer and constructor choices; and
    (iv)~greater recovery of keys and foreign keys when structural evidence is available
    (\S\ref{sec:results}).

\end{itemize}

\section{Related Work}
\label{sec:related_work}

We position \textsf{Text2Schema}, \textsf{Schema2Text}, and the Schema$\rightarrow$Text$\rightarrow$ {Schema} round trip against four lines of related work: foundations of relational schema design, schema construction from natural language requirements, natural language interfaces over existing schemata, and LLM-based relational schema construction.

\spara{Foundations of Relational Schema Design.}
Relational schema design is a foundational problem in database systems. Codd~\cite{codd1970relational} introduced the relational model, establishing relations as a logical abstraction of structured data independent of their physical storage and access paths. Chen~\cite{chen1976entity} complemented this relational view with the entity-relationship model, which explicitly models the semantics through entities, relationships, attributes, and structural constraints, providing a systematic bridge between application requirements and logical database design. Bernstein~\cite{bernstein1976synthesizing} subsequently formalized schema synthesis from functional dependencies (FDs), establishing dependencies as a formal basis for constructing normalized relational schemata. Later work automated parts of this process by discovering  dependencies from existing relational instances and using them to support normalization and schema restructuring ~\cite{papenbrock2015functional, papenbrock2017data}. These approaches provide formal mechanisms for deriving or refining relational structure but generally assume that the relevant data, dependencies, or schema elements are already available.

\spara{Schema Construction from Natural Language Requirements.}
A complementary line of work derives database designs from textual requirements. Early requirements-engineering approaches combine linguistic analysis, dependency parsing, hand-crafted extraction rules, and ontology-driven reasoning to identify entities, attributes, and relationships in natural language specifications~\cite{deeptimahanti2011semi, lucassen2017extracting}. They show that conceptual structure can be recovered from text, but they target conceptual models such as UML or entity relationship diagrams and depend on manually designed rules or task-specific extraction pipelines. As a result, they do not evaluate the generation of executable relational schemata, including tables, attributes, domains, keys, and referential constraints. \textsf{\bfseries Text2Schema} extends this direction by assessing whether LLMs can construct these relational components directly from natural language requirements.

\spara{Natural Language Interfaces over Existing Schemata.}
Advancements in LLMs have enabled substantial progress in translation between natural language and SQL. In \textsf{Text2SQL}, approaches such as DIN-SQL~\cite{pourreza2023dinsql} employ task decomposition and self-correction, while DAIL-SQL~\cite{gao2024text} investigates prompt engineering and in-context learning for generating SQL queries from natural language questions over existing relational schemata. Benchmarks such as WikiSQL~\cite{zhong2017seq2sql}, Spider~\cite{yu2018spider}, and
BIRD~\cite{li2023can} have supported progress in this direction. However, these approaches assume that the underlying relational schema is already available. They evaluate an LLM's ability to generate queries over a given schema, rather than its ability to construct or faithfully
represent the schema itself.

The complementary direction, \textsf{SQL2Text}, translates SQL queries into natural language descriptions to support explainability and query verification~\cite{xu2018sql, camara2024large, chakrabarti2025sql, al2025semantic}. Although it verbalizes a formal database artifact, the object being described is an individual query rather than the schema. Consequently, neither \textsf{Text2SQL} nor \textsf{SQL2Text} evaluates whether schema-level information (tables, attributes, domains, keys, and referential dependencies) survives translation between formal and natural language representations. Our work takes the relational schema itself as the object of translation and evaluation.

\spara{LLM-Based Schema Construction.}
Recent work has started to apply LLMs directly to relational schema
construction. SchemaAgent~\cite{wang2025schemaagent} formulates
\textsf{Text2Schema} as generating a relational schema from natural language requirements, using a multi-agent framework with stages for requirement analysis, conceptual modeling, logical design, and validation. Its evaluation covers relations, attributes, data types, primary keys, and foreign keys, making it the closest prior work to our \textsf{Text2Schema} setting.

In the complementary direction, Avignone et al.~\cite{avignone2025exploring} investigate LLMs' ability to generate natural language descriptions of entity-relationship (ER) conceptual models. They evaluate the verbalization of ER constructs using
different LLM backbones and prompting strategies. While this is closely related to our \textsf{Schema2Text} setting, their evaluation focuses on conceptual ER models rather than relational DDL and does not examine whether generated descriptions preserve sufficient structural information
for schema reconstruction.

SchemaBench~\cite{chafik_2026_schemabench} provides pairs of natural language descriptions and their corresponding relational schemata, supporting the evaluation of requirement-to-schema generation.
Although it serves as a dataset for \textsf{Text2Schema}, it does not provide a unified evaluation of schema verbalization and round-trip structural preservation. We utilize SchemaBench as one of the three
schema collections in our evaluation.

These experiments examine complementary aspects of schema construction and verbalization but do not jointly evaluate the preservation of relational structure across both directions. To the best of our knowledge, \textsf{SchemaBridge} is the first unified evaluation
framework to assess schema design, schema comprehension, and Schema$\rightarrow$Text$\rightarrow$Schema round-trip preservation,
while evaluating structural information loss across verbalization and reconstruction stages.

\section{SchemaBridge: A Testbed for Schema Design and Comprehension}
\label{sec:methodology}

\begin{figure}[t]
  \centering
   \includegraphics[width=\linewidth]{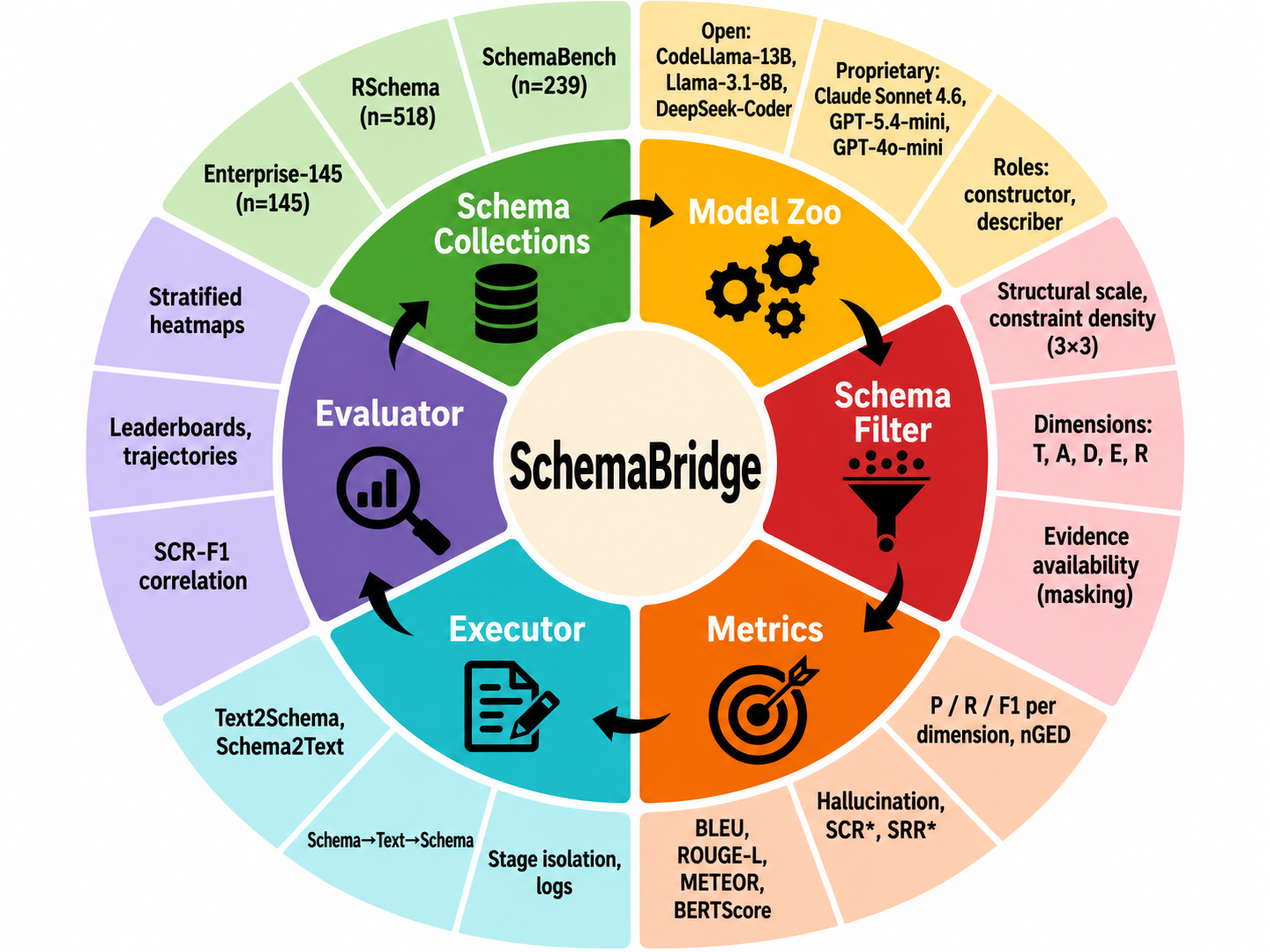}
  \caption{\textsf{SchemaBridge} execution and evaluation workflow. Schemata from three collections are stratified by size and constraint complexity, passed through the model zoo in the roles of describer and constructor, and scored with element-level, schema-level, and description-level metrics. Masking is applied to test structural inferability. }
  \label{fig:schemabridge_overview}
\end{figure}

Figure~\ref{fig:schemabridge_overview} provides an overview of the SchemaBridge execution and evaluation workflow. The following subsections describe its key components in detail.

\subsection{Evaluation Scope}
\label{sec:evaluation_scope}

SchemaBridge evaluates three capabilities over a common relational schema
representation:

\begin{itemize}
  \item \textsf{Text2Schema} (design): generate a schema from a natural language requirement.
  \item \textsf{Schema2Text} (comprehension): describe an existing schema in natural language.
  \item Schema$\rightarrow$Text$\rightarrow$Schema
  (preservation): reconstruct a schema from its own description.
\end{itemize}

We regard a description as faithful if it preserves schema-critical information, and we measure this by how much of the schema can be recovered from it, rather than by its textual similarity to a reference.

Outputs are analyzed at three levels: the \emph{schema} level (divergence from the gold schema), the \emph{dimension} level (fidelity per structural dimension), and the \emph{scenario} level (aggregation over models, datasets, prompts, and schema complexity).

\spara{Out of scope.} We evaluate schema structure only. We do not assess query generation, instance data, normal forms or functional dependencies, referential actions, default values, or \texttt{CHECK} constraints. Constraints beyond keys and foreign keys, such as functional dependencies and cardinality bounds, raise separate questions of runtime enforcement \cite{kalumin2026}, which we leave for future work.

\subsection{Schema Content and Structural Fidelity}
\label{sec:fidelity}

\paragraph{Identifier and type normalization}
Identifiers are normalized by lowercasing and removing quoting characters, spaces, underscores, and hyphens. SQL data types are compared using compatible integer, string, and binary families, with normalized base-type matching otherwise. Domain fidelity additionally requires matching nullability. Constraint names, engine clauses, and SQL keyword case are ignored. Our evaluation uses strict reference-based matching rather than functional equivalence.

\paragraph{Schema content}
The content of a schema $S$ is a tuple of five sets
$\mathcal{C}(S)=(\mathcal{C}_T,\mathcal{C}_A,\mathcal{C}_D,\mathcal{C}_E,\mathcal{C}_R)$,
one per structural dimension, with dimension set
$\mathbb{D}=\{T,A,D,E,R\}$:
\begin{itemize}
  \item \textbf{Tables} ($\mathcal{C}_T$): the normalized names of the tables.
  \item \textbf{Attributes} ($\mathcal{C}_A$): the $(t,c)$ pairs of table $t$ and column $c$, capturing column-level structure beyond the set of relations.
  \item \textbf{Domain integrity} ($\mathcal{C}_D$): tuples $(t,c,\tau,\nu)$ of table, column, data type $\tau$, and nullability
  $\nu\in\{\texttt{NULL},\texttt{NOT NULL}\}$. A column is correct only if both $\tau$ and $\nu$ match. We score only these two properties, not the full range of domain constraints (e.g., defaults, \texttt{CHECK}).
  \item \textbf{Entity integrity} ($\mathcal{C}_E$): triples $(t,K,\kappa)$ of table $t$, key columns $K$ (single or composite), and constraint type $\kappa\in\{\texttt{PK},\texttt{UNIQUE}\}$.
  \item \textbf{Referential integrity} ($\mathcal{C}_R$): foreign keys as
  directed 4-tuples $(t_s,A_s,t_t,A_t)$ of source table, referencing
  columns, target table, and referenced columns. A foreign key is correct
  only if all four components match. Referential actions
  (\texttt{ON DELETE}/\texttt{ON UPDATE}) are not scored.
\end{itemize}

\begin{figure}[t]
    \centering
    \includegraphics[width=0.98\columnwidth]{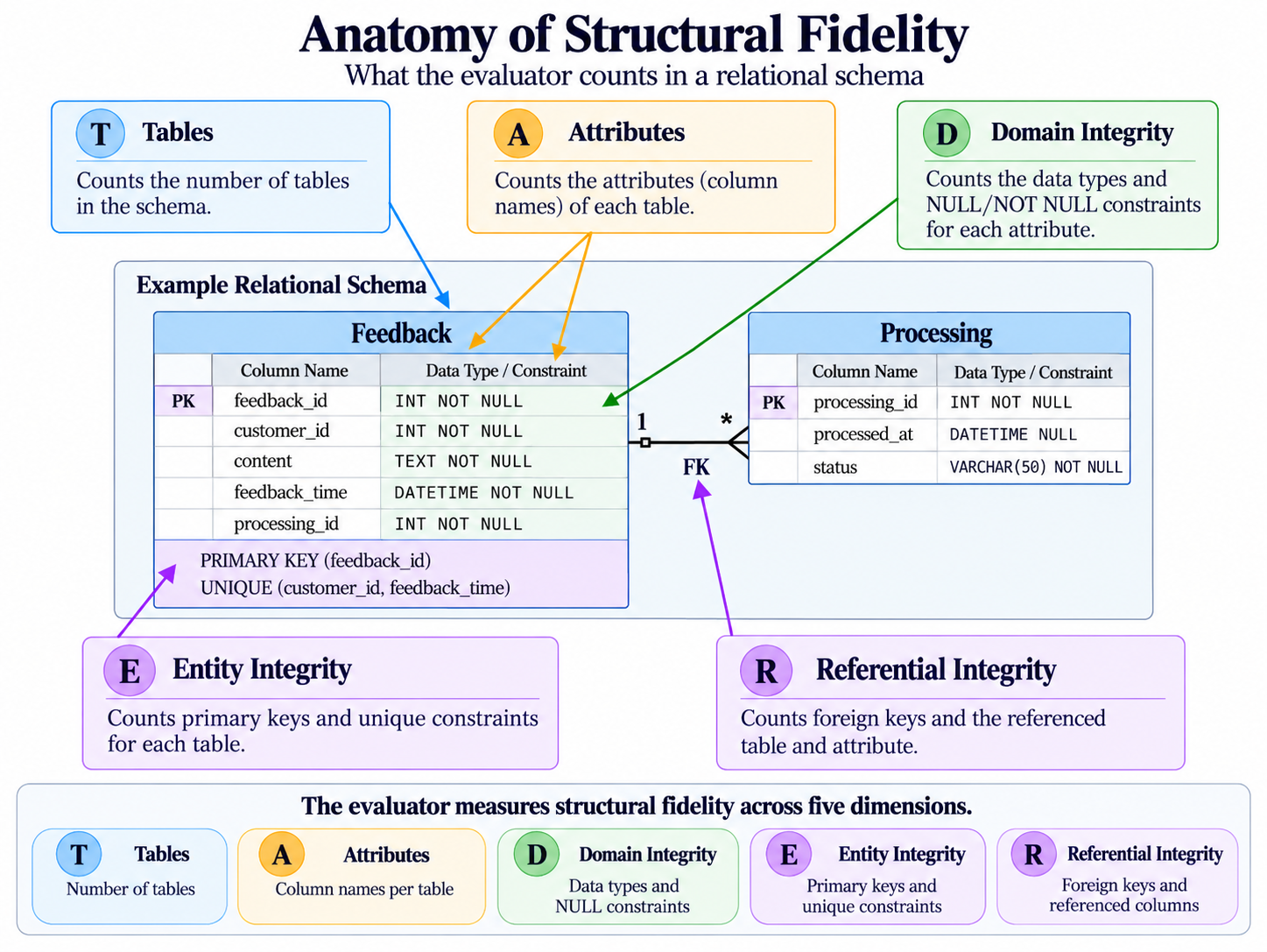}
    \caption{Anatomy of structural fidelity. The evaluator measures schema fidelity across five dimensions:
    Tables, Attributes, Domain integrity, Entity integrity, and Referential integrity. Using a small two-table example, the figure highlights exactly which schema elements are counted in each dimension.}
    \label{fig:structural-fidelity}
\end{figure}

Two schemata are \emph{content-equivalent} when all five sets coincide after normalization. \emph{Structural fidelity} is the agreement between a reconstructed schema $\hat S$ and the gold schema $S$ on each component of $\mathcal{C}(\cdot)$, reported separately per dimension (\S\ref{sec:evaluation_metrics}). Figure~\ref{fig:structural-fidelity} illustrates which schema elements each dimension counts.

\subsection{Tasks and Round-Trip Diagnosis}
\label{sec:tasks}

We evaluate three complementary tasks using a gold relational schema $S$ and its natural language requirements $Y$. Let $\gamma$ denote an LLM-based schema constructor, $\delta$ a schema describer, $T$ the generated natural language description, and $\hat{S}$ the reconstructed schema.

The three tasks are defined as:
\[
\begin{gathered}
\textsf{Text2Schema}:~\hat S=\gamma(Y),\qquad
\textsf{Schema2Text}:~T=\delta(S),\\
\textsf{RoundTrip}:~\hat S=\gamma(\delta(S)).
\end{gathered}
\]

In \textsf{Text2Schema}, the constructed schema $\hat{S}$ is compared against the gold schema $S$. In \textsf{Schema2Text}, the describer generates a natural language description $T$ of $S$. The Schema$\rightarrow$Text$\rightarrow$Schema round trip combines both transformations: $S \xrightarrow{\ \delta\ } T
  \xrightarrow{\ \gamma\ } \hat{S}$.

The reconstructed schema $\hat{S}$ is evaluated against the original schema $S$ using structural fidelity measures in \S\ref{sec:evaluation_metrics}, without requiring a gold natural language description.

To diagnose information loss, we first hold the constructor $\gamma$ fixed while varying the describer $\delta$, thereby measuring the effect of schema verbalization. Conversely, we hold the generated description $T$ fixed while varying the constructor $\gamma$, thereby measuring differences in reconstruction performance. These controlled comparisons distinguish the contributions of the two stages to overall structural fidelity.

\subsection{Masking and Inferability}
\label{sec:masking}
To assess structural inferability, we mask one schema dimension at a time. For $d\in\mathbb{D}$, the masked view $S^{\setminus d}$ hides the corresponding schema information while retaining the remaining structure. For $M_D$, we retain the historical masking protocol, which hides data types, nullability, defaults, and auto-increment properties. However, domain recovery is evaluated only through joint data-type family and nullability matching, consistent with
\S\ref{sec:fidelity}.

Hidden elements are classified as \emph{inferable} when the remaining schema provides heuristic evidence
for recovery, such as identifier patterns or visible relationships; otherwise, they are \emph{non-inferable}. These labels reflect structural clues rather than logical entailment. Round-trip retention is measured as the ratio of Round-trip to Direct recall on paired masked elements.

\subsection{Evaluation Metrics}
\label{sec:evaluation_metrics}

Metrics operate at three levels (Table~\ref{tab:metrics}). Element-level overlap and schema-level distance compare a reconstructed schema $\hat S$ with the gold schema $S$ and apply to \textsf{Text2Schema} and the round trip; description-level metrics apply to \textsf{Schema2Text}. Failed or unparseable generations are treated as empty predictions over the complete evaluation set. They receive zero structural $F_1$ and nGED$=1$; failure counts are reported separately.

\paragraph{Element-level overlap}
For each dimension $d\in\mathbb{D}$, let $G_d=\mathcal{C}_d(S)$,
$\hat G_d=\mathcal{C}_d(\hat S)$, and $M_d=G_d\cap\hat G_d$. Precision,
recall, and F$_1$~\cite{cleverdon1967cranfield,rijsbergen1979information} are
\[
P_d=\frac{|M_d|}{|\hat G_d|},\qquad
R_d=\frac{|M_d|}{|G_d|},\qquad
F_{1,d}=\frac{2P_dR_d}{P_d+R_d},
\]

with any ratio set to $0$ when its denominator is $0$. Precision measures specificity (how much of the output is grounded in the gold schema) and recall measures coverage (how much of the gold schema is recovered). We report micro-averages by aggregating matched, predicted, and gold counts before computing each ratio.

\paragraph{Hallucination}
We define hallucination as reference-relative structural over-generation. For each dimension $d\in\mathbb{D}$,
the hallucination rate is

\[
H=\frac{1}{4}\sum_{d\in\{T,A,E,R\}}
\frac{\max(0,\sum_i P_{i,d}-\sum_i M_{i,d})}
{\sum_i P_{i,d}},
\]

\noindent where $P_{i,d}$ and $M_{i,d}$ are the predicted and
matched element counts. A zero denominator is considered
zero. Failed generations contribute empty predictions.
Strict matching may count valid alternative designs as
false positives; thus, hallucination measures reference
disagreement rather than semantic invalidity.


\paragraph{Schema-level distance (nGED)} To summarize divergence in a single number, we use a unit-cost structural
distance over three element sets derived from $\mathcal{C}(\cdot)$:
tables $\mathcal{T}=\mathcal{C}_T$, attribute labels
$\mathcal{A}=\{(t,c,\mathit{family}(\tau),\nu)\}$ (name, type family, and
nullability), and foreign keys $\mathcal{F}=\mathcal{C}_R$. With
$\hat{\mathcal{T}},\hat{\mathcal{A}},\hat{\mathcal{F}}$ defined likewise
for $\hat S$,

{\footnotesize
\[
\mathrm{nGED}(S,\hat S)=
\frac{\sum_{X\in\{\mathcal{T},\mathcal{A},\mathcal{F}\}}|X\,\triangle\,\hat X|}
     {\max\!\big(|\mathcal{T}|+|\mathcal{A}|+|\mathcal{F}|
        +|\hat{\mathcal{T}}|+|\hat{\mathcal{A}}|+|\hat{\mathcal{F}}|,\,1\big)}
\in[0,1].
\]
}

Equivalently, it is a simplified graph edit distance~\cite{riesen2009approximate} in which tables are nodes, attribute labels are node labels, and foreign keys are directed edges. Lower is better; $0$ means identical element sets. We report the mean nGED over schemata. Since an attribute label includes type family and nullability, one type mismatch costs two units (one missing, one extra label), so nGED is sensitive to domain errors. In E0, we use a type-independent variant of nGED, representing attributes by their table-qualified names $(t,c)$, to ensure comparability with untyped reference and baseline schemata.

\paragraph{Description quality (Schema2Text)}
We compare a generated description $T$ with the gold description using
BLEU-4~\cite{papineni2002bleu}, ROUGE-L~\cite{lin2004rouge}, and
BERTScore-F$_1$~\cite{zhang2019bertscore}. These measure lexical or semantic similarity to the reference, not whether $T$ contains the schema's content. To measure content directly, we define \emph{Schema Coverage Rate} (SCR), the fraction of source table and column identifiers mentioned in $T$:

\[
\mathrm{SCR}(T,S)=\frac{|\{e\in E(S):\mathrm{norm}(e)\ \text{occurs in}\ \mathrm{norm}(T)\}|}{|E(S)|},
\]
\[
E(S)=\mathrm{Tables}(S)\cup\mathrm{Columns}(S),
\]

\noindent where $\mathrm{norm}$ is the identifier normalization of
\S\ref{sec:fidelity}. We report the macro-average over schemata. SCR does not assess data types, keys, foreign keys, or other constraints.

\begin{table}[t]
\centering
\caption{Evaluation metrics.}
\label{tab:metrics}
\footnotesize
\begin{tabular}{@{}llll@{}}
\toprule
Level & Metric & Compared against & Better \\
\midrule
Element & P/R/F1 per dimension & gold schema & $\uparrow$ \\
Element & Hallucination rate & gold schema & $\downarrow$ \\
Schema  & nGED & gold schema & $\downarrow$ \\
Description & BLEU-4, ROUGE-L, BERTScore & gold description & $\uparrow$ \\
Description & SCR & source schema & $\uparrow$ \\
\bottomrule
\end{tabular}
\end{table}

\section{Experimental Setup}
\label{sec:experimental_setup}

\subsection{Datasets}
\label{sec:datasets}

We evaluate \textsf{SchemaBridge} on three complementary schema collections: \emph{\textit{Enterprise-145}}~\cite{ctuprague}, \emph{\textit{RSchema}}~\cite{wang2025schemaagent}, and \emph{\textit{SchemaBench}}~\cite{chafik_2026_schemabench}, which together span enterprise-scale DDL, real-world relational designs, and software-project schemata paired with natural language specifications. This diversity lets us test whether LLM behavior is consistent across schema sources, sizes, domains, and levels of natural language supervision. Enterprise-145 and SchemaBench provide DDL references, while RSchema provides 518 logical annotations and 381 matched DDL references. Experiments requiring DDL use the prepared schema representations
Gold DDL schemata are used as input for \textsf{Schema2Text} and the \textsf{Schema}$\to$\textsf{ Text}$\to$\textsf{Schema} round trip, in which the gold DDL is verbalized and then reconstructed, and structural preservation is measured against the original. Only the collections providing natural language requirements (\emph{\textit{RSchema}} and \emph{\textit{SchemaBench}}) are used for \textsf{Text2Schema}, where we test whether LLMs can generate executable DDL from textual requirements.

\textbf{\em \textit{Enterprise-145}.} Our collection comprises 145 enterprise-grade relational schemata drawn from publicly available database benchmarks and real-world data integration repositories, spanning e-commerce, healthcare, logistics, finance, and academic information systems. Each schema is retained in its original DDL form, preserving column-level metadata including data types, \texttt{NOT NULL} flags, primary keys, unique constraints, and foreign keys. These DDL definitions serve as the gold schema $S$ in our evaluation. However, \emph{\textit{Enterprise-145}} does not provide human-authored natural language descriptions of its schema. Therefore, we use it for \textsf{Schema2Text} and Schema$\to$Text$\to$Schema evaluation. Its larger and more densely interconnected schemata provide a challenging setting for evaluating structural preservation.

\textbf{\em RSchema.}
We use all 518 \textit{RSchema} logical references for structural evaluation and 381 matched DDL references for domain integrity~\cite{wang2025schemaagent}. The logical annotations do not specify SQL data types or nullability; therefore, domain fidelity is evaluated only on the DDL-backed subset. \textit{RSchema} contains natural language requirements paired with annotated relational schemata. Its source material includes DDL from SchemaPile, constructed industry scenarios, and publicly available database-design examples. Some requirements were generated using LLMs, while the dataset construction process involved manual refinement, cross-review, and expert validation. The reference logical schemata specify tables, attributes, primary keys, and foreign keys. We use these annotations as the gold reference and apply the identifier and type normalization described in \S\ref{sec:fidelity}. In our masking experiments, table names are replaced with placeholders to assess structural inferability. This experimental masking is distinct from anonymization of the original dataset.

\textbf{\em SchemaBench.} We use \emph{SchemaBench}, a collection of 239 project-to-schema pairs derived from real open-source applications and public schema-design repositories. The pairs span 27 application categories, including content management, SaaS, subscription systems, IT asset management, and social media. Each instance consists of a natural-language project specification, a corresponding SQL DDL schema, and metadata identifying the schema's source. We treat the released DDL as the gold reference. The dataset documentation states that the examples were manually curated and semi-automatically validated, but it does not describe the validation procedure or indicate whether the specifications were written entirely by humans or with LLM assistance. We therefore treat the provided descriptions as given requirements and evaluate structural fidelity against the reference schemas.

\begin{table*}[t]
\centering
\caption{Statistics of the three schema collections.
Means are computed per schema. Composite FKs (\%)
denotes the proportion of foreign-key constraints
involving multiple columns. UNIQUE (\%) denotes
the proportion of schemata containing at least one
UNIQUE constraint, excluding primary keys.}
\label{tab:dataset_statistics}

\footnotesize
\setlength{\tabcolsep}{4pt}
\renewcommand{\arraystretch}{1.12}

\begin{tabular*}{\textwidth}{@{\extracolsep{\fill}}lrrrrrrll@{}}
\toprule
\textbf{Collection}
& \textbf{\# Schemata}
& \textbf{Tables}
& \textbf{Columns}
& \textbf{FKs}
& \textbf{Composite FKs (\%)}
& \textbf{UNIQUE (\%)}
& \textbf{SQL Dialect}
& \textbf{Gold NL} \\
\midrule
\emph{Enterprise-145}
& 145 & 13.9 & 119.2 & 13.4 & 4.2
& 15.9 & MySQL & No \\

\emph{RSchema}
& 518 & 4.7 & 22.6 & 4.1 & 0.1
& 0.0 & DBMS-agnostic & Yes \\

\emph{SchemaBench}
& 239 & 11.0 & 79.2 & 11.8 & 2.4
& 11.3 & MySQL-style & Yes \\
\bottomrule
\end{tabular*}

\end{table*}

Table~\ref{tab:dataset_statistics} summarizes the structural characteristics of the three datasets. \emph{\textit{Enterprise-145}} contains the largest schemata on average, with 13.9 tables, 119.2 columns, and 13.4 foreign keys per schema, followed by \emph{\textit{SchemaBench}} with 11.0 tables, 79.2 columns, and 11.8 foreign keys. In contrast, \emph{\textit{RSchema}} contains smaller schemata, averaging 4.7 tables and 22.6 columns, with few composite foreign keys and no UNIQUE constraints. These differences motivate our stratification by schema size and constraint complexity.

\subsection{Dataset Stratification}
\label{sec:schema_strata}

\begin{figure}[t]
    \centering
    \includegraphics[width=0.93\columnwidth]{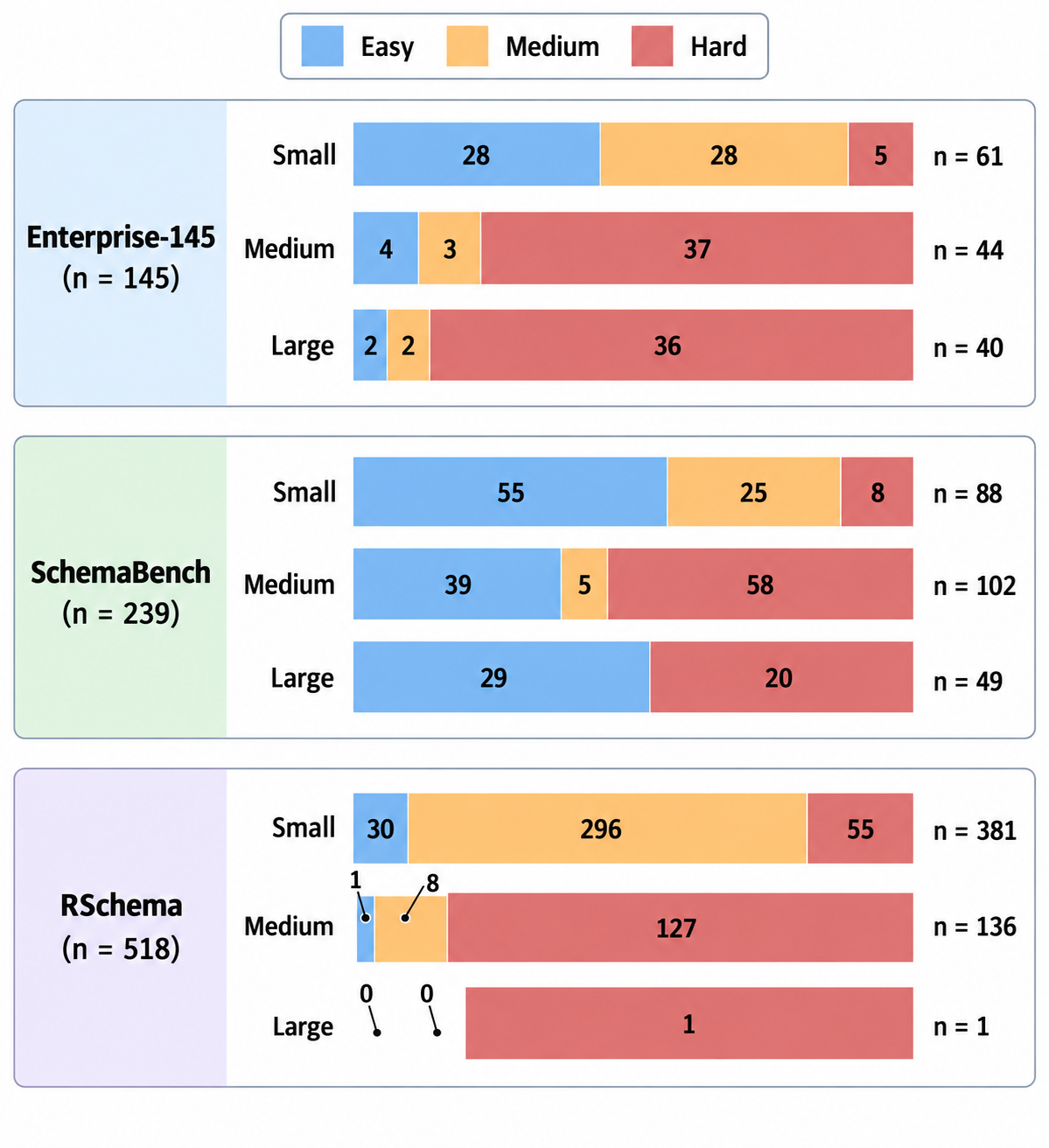}
    \caption{Distribution of 902 schemata by size and constraint complexity across three datasets. Bars show category proportions, segment labels indicate counts, and $n$ denotes tier totals.}
    \label{fig:dataset-landscape}
\end{figure}

We stratify each gold schema along two complementary dimensions: \textsc{Size} and \textsc{Constraint Complexity}. \textsc{Size} captures the overall schema scale, measured by the number of tables and columns. \textsc{Constraint Complexity} captures the difficulty of preserving relational constraints, considering foreign keys, composite foreign keys, UNIQUE constraints, and table width. These differences enable us to evaluate schema scale and constraint complexity separately.

\noindent The content score is
\begingroup
\setlength{\abovedisplayskip}{2pt}
\setlength{\belowdisplayskip}{2pt}
\setlength{\abovedisplayshortskip}{2pt}
\setlength{\belowdisplayshortskip}{2pt}
\[
s=2\min(f,10)+3c+\min(u,5)
+\tfrac15\max\!\left(0,\tfrac{n_{\mathrm{col}}}{n_{\mathrm{tab}}}-10\right),
\]
\endgroup

$f$ denotes the number of foreign keys, $c$ the number of composite foreign keys, $u$ the number of UNIQUE constraints, and $n_{\mathrm{col}}/n_{\mathrm{tab}}$ the average number of columns per table.

The weights reflect the relative complexity of relational constraints. Foreign keys receive a weight of 2 as they encode inter-table dependencies, while composite foreign keys receive an additional weight of 3 to account for multi-column relationships. UNIQUE constraints receive a weight of 1, as they capture table-level constraints. The smaller weight of $1/5$ accounts for wider tables without dominating the score. By limiting foreign keys and UNIQUE contributions, it limits the influence of schema size, which is evaluated separately.

These two axes define a $3 \times 3$ evaluation matrix (Table~\ref{tab:matrix-rules}). All stratification features are computed exclusively from the gold schema.

\begin{table}[t]
\centering
\caption{Classification rules.}
\label{tab:matrix-rules}
\footnotesize
\renewcommand{\arraystretch}{0.9}
\setlength{\tabcolsep}{3pt}
\begin{tabular*}{\columnwidth}{@{\extracolsep{\fill}}lll@{}}
\toprule
\textbf{Dimension} & \textbf{Category} & \textbf{Criterion} \\
\midrule
\multirow{3}{*}{Size}
 & Small  & $\leq 5$ tables, $\leq 50$ columns \\
 & Medium & $\leq 15$ tables, $\leq 150$ columns \\
 & Large  & Otherwise \\
\midrule
\multirow{3}{*}{Content}
 & Easy   & $s < 4$ \\
 & Medium & $4 \leq s < 10$ \\
 & Hard   & $s \geq 10$ \\
\bottomrule
\end{tabular*}
\end{table}

\noindent\textbf{Threshold Selection.}
We adopt heuristic size thresholds of 5 and 15 tables,
paired with 50 and 150 columns, respectively, assuming
approximately 10 columns per table. For structural
complexity, the thresholds $s=4$ and $s=10$ correspond
to two and five ordinary foreign keys, respectively,
when other score components are zero. These thresholds
provide practical boundaries for stratified evaluation
rather than universal measures of schema difficulty.

Note that \textsc{Easy} does not imply easier reconstruction: constraint-sparse schemata leave an LLM more room to add structure that the gold design omits, which exact matching counts as errors (\S\ref{sec:e0} and \ref{sec:e4}).

\subsection{Model Zoo.} 
We use six LLMs to evaluate. The open-weight models, accessed through the HuggingFace Inference API, are the code-specialized CodeLlama-13B~\cite{roziere2023code} and DeepSeek-Coder-6.7B~\cite{guo2024deepseekcoder}, and the general-purpose Llama-3.1-8B~\cite{grattafiori2024llama}. The proprietary instruction-tuned models, accessed through their providers' APIs, are Claude Sonnet~4.6~\cite{anthropic2026claudesonnet46}, GPT-5.4-mini~\cite{openai2026gpt54mini}, and GPT-4o-mini~\cite{openai2024gpt4omini}. This selection spans model families, code specialization, and scale, which lets us examine whether relational schema reasoning varies along these axes. The first five models serve as both describers and constructors; GPT-4o-mini serves only as the fixed constructor for the round trip (\S\ref{sec:e2}) and for the direct masking track (\S\ref{sec:e4}).

\subsection{Prompting Protocol}
\label{sec:prompting_protocol}

We use a fixed constructor prompt that requests MySQL-style DDL and structured metadata in JSON, and two describer prompts: basic, requesting a concise summary, and detailed, explicitly requesting tables, columns, types, nullability, keys, and relationships. Complete templates are released
with the reproducibility artifacts.

\paragraph{Constructor}
A single fixed prompt asks the model to convert a natural language description, either a gold requirement or a generated description, into one JSON object with two fields: \texttt{ddl}, containing MySQL-style table definitions and constraints, and \texttt{metadata}, recording the database name, table names, and column properties. The prompt instructs the model to choose names consistent with the description and to add cascading update and delete actions where logical. The same prompt is used for all constructor models and for round-trip reconstruction.

\begin{lstlisting}[
  style=promptstyle,
  breaklines=true,
  breakindent=0pt,
  postbreak={},
  columns=fullflexible,
  xleftmargin=0pt,
  xrightmargin=0pt,
  caption={Basic describer prompt (verbatim).},
  label={lst:basic_describer_prompt},
  captionpos=t,
  abovecaptionskip=0pt,
  belowcaptionskip=0pt,
  aboveskip=2pt,
  belowskip=3pt
]
Summarize this database schema in one short paragraph of plain English: what it is for, main tables, and how they relate. Human-readable, no code.
\end{lstlisting}

\paragraph{Basic describer}
The basic prompt (Listing~\ref{lst:basic_describer_prompt}) requests a short, human-readable summary and does not require exhaustive structural coverage.

\paragraph{Detailed describer}
The detailed prompt keeps the single-paragraph format but explicitly requests the schema-critical information. Listing~\ref{lst:detailed_describer_prompt} shows the excerpt on constraint coverage. Unlike the basic
prompt, it does not require the paragraph to be short.

\begin{lstlisting}[
  style=promptstyle,
  breaklines=true,
  breakindent=0pt,
  postbreak={},
  columns=fullflexible,
  xleftmargin=0pt,
  xrightmargin=0pt,
  caption={Detailed describer prompt (verbatim excerpt).},
  label={lst:detailed_describer_prompt},
  captionpos=t,
  abovecaptionskip=0pt,
  belowcaptionskip=0pt,
  aboveskip=2pt,
  belowskip=3pt
]
primary keys and any composite or unique constraints (entity integrity); every foreign key and which columns reference which tables, plus update and delete rules like CASCADE or RESTRICT (referential integrity); and any other indexes or constraints. Be exhaustive but write it all as one continuous, readable paragraph that a non-expert can follow.
\end{lstlisting}

\paragraph{Reproducibility}
The complete, unabridged constructor and describer templates, including their output specifications,
are provided in the reproducibility package: \url{https://github.com/iamsriom/SchemaBridge}

\section{Experimental Evaluation}
\label{sec:results}

\subsection{Research Questions and Protocol}
\label{sec:rqs}

We organize the evaluation around four research questions.

\begin{itemize}
\item[\textbf{RQ1}] \emph{Design.} How accurately do LLMs construct relational schemata from natural language requirements, and how do prompting strategies affect structural fidelity?

\item[\textbf{RQ2}] \emph{Comprehension.} Can LLMs verbalize an existing schema without losing content, and do text-similarity metrics detect any loss?

\item[\textbf{RQ3}] \emph{Preservation.} How much schema content survives the Schema$\to$Text$\to$Schema round trip, and does the loss arise in the describer or in the constructor?

\item[\textbf{RQ4}] \emph{Structural inferability.} When schema content is hidden, which elements can be recovered from the remaining relational evidence, and which are supplied mainly by learned design conventions?
\end{itemize}

\begin{table}[t]
\centering
\caption{Experiments and the research questions they address.}
\label{tab:rq_map}
\scriptsize
\setlength{\tabcolsep}{3pt}
\renewcommand{\arraystretch}{1.0}
\resizebox{\columnwidth}{!}{%
\begin{tabular}{@{}lccccc@{}}
\toprule
\textbf{Experiment}
& \textbf{RQ1}
& \textbf{RQ2}
& \textbf{RQ3}
& \textbf{RQ4}
& \textbf{Sec.} \\
\midrule

E0 Text2Schema from gold NL
    & \cmark & & & & \ref{sec:e0} \\

E1 Schema2Text verbalization
    & & \cmark & & & \ref{sec:e1} \\

E2 Round-trip preservation
    & & \cmark & \cmark & & \ref{sec:e2} \\

E3 Constructor isolation
    & & & \cmark & & \ref{sec:e3} \\

E4 Masking and inferability
    & & & \cmark & \cmark & \ref{sec:e4} \\

E5 Scalability and cost
    & \cmark & & & & \ref{sec:e5} \\


\bottomrule
\end{tabular}%
}
\end{table}


\subsection{E0: Schema Design from Natural Language Requirements}
\label{sec:e0}

\paragraph{Setup}
Each of the five constructors receives dataset-provided requirements for 518 {\em \textit{RSchema}} and 239 {\em \textit{SchemaBench}} schemata using the fixed constructor prompt (\S\ref{sec:prompting_protocol}). Predictions are evaluated against gold references: all 518 logical {\em \textit{RSchema}} references for tables, attributes, entity, and referential integrity; 381 matched DDL references for domain integrity; and original DDL references for {\em \textit{SchemaBench}}. We report micro-averaged $F_1$, four-dimensional hallucination (excluding domain) and type-independent nGED. Failed or unparseable generations are treated as empty predictions (nGED $=1$ for nonempty gold), without an additional hallucination penalty.

\begin{table*}[t]
\centering
\small
\setlength{\tabcolsep}{3pt}
\renewcommand{\arraystretch}{1.0}
\caption{E0: Direct Text2Schema from gold natural-language requirements.
Mean structural nGED and four-dimensional hallucination rate (lower is better),
and micro-averaged $F_1$ per dimension (higher is better).
\emph{RSchema} uses 518 logical references for structural evaluation and 381
matched genuine DDL references for domain $F_1$. Best values per dataset and
column are in bold.}
\label{tab:e0_overall}
\begin{tabular*}{\textwidth}{@{\extracolsep{\fill}}llccccccc@{}}
\toprule
\multirow{2}{*}{\textbf{Dataset}}
& \multirow{2}{*}{\textbf{Constructor}}
& \multirow{2}{*}{\textbf{nGED} $\downarrow$}
& \multirow{2}{*}{\textbf{Hall.} $\downarrow$}
& \multicolumn{5}{c}{\textbf{Structural} $F_1$ $\uparrow$} \\
\cmidrule(lr){5-9}
& & & & \textbf{Tables} & \textbf{Attr.} & \textbf{Domain}
& \textbf{Entity} & \textbf{Ref.} \\
\midrule
\multirow{5}{*}{\emph{RSchema}}
& CodeLlama-13B & 0.583 & 0.669 & 0.554 & 0.414 & 0.115 & 0.301 & 0.141 \\
& Llama-3.1-8B & 0.615 & 0.696 & 0.572 & 0.377 & 0.121 & 0.199 & 0.104 \\
& DeepSeek-Coder-6.7B & 0.561 & \textbf{0.595} & 0.566 & \textbf{0.443} & 0.150 & 0.391 & 0.184 \\
& Claude Sonnet 4.6 & 0.560 & 0.635 & \textbf{0.602} & 0.401 & 0.208 & \textbf{0.407} & 0.204 \\
& GPT-5.4-mini & \textbf{0.553} & 0.605 & 0.600 & 0.433 & \textbf{0.223} & 0.399 & \textbf{0.225} \\
\midrule
\multirow{5}{*}{\emph{SchemaBench}}
& CodeLlama-13B & 0.794 & 0.773 & 0.350 & 0.125 & 0.067 & 0.183 & 0.082 \\
& Llama-3.1-8B & 0.810 & 0.768 & 0.363 & 0.112 & 0.061 & 0.145 & 0.063 \\
& DeepSeek-Coder-6.7B & 0.731 & 0.641 & 0.398 & 0.185 & 0.100 & 0.240 & 0.128 \\
& Claude Sonnet 4.6 & \textbf{0.574} & \textbf{0.598} & \textbf{0.530} & \textbf{0.358} & \textbf{0.247} & \textbf{0.345} & \textbf{0.263} \\
& GPT-5.4-mini & 0.675 & 0.651 & 0.465 & 0.245 & 0.159 & 0.255 & 0.174 \\
\bottomrule
\end{tabular*}
\end{table*}

\paragraph{Requirements under-specify schema content}
Table~\ref{tab:e0_overall} shows that models could recover tables more reliably than domain and referential constraints. Natural language requirements often describe application entities but leave data types, nullability, and relationships implicit. Across both datasets, the comparatively low domain and referential $F_1$ values are consistent with requirements that specify entities more explicitly than attribute types, nullability, and foreign-key constraints.

\spara{Baselines and scoring.}
We compare Direct construction with one-shot, one-shot Chain-of-Thought
(CoT), and few-shot prompting configurations based on
SchemaAgent~\cite{wang2025schemaagent}, using three open-weight LLMs
(Table~\ref{tab:e0_baselines}). We exclude the complete SchemaAgent
multi-agent framework because its released implementation could not be
reproduced in our environment. All outputs are evaluated under the same reference-based structural metrics. Domain $F_1$ is omitted for untyped baseline outputs.

\begin{table*}[t]
\centering
\small
\setlength{\tabcolsep}{3pt}
\renewcommand{\arraystretch}{1.0}
\caption{E0: Text2Schema performance of open-weight LLMs
on \emph{\textit{RSchema}}, measured by structural nGED,
four-dimensional hallucination, and micro-averaged
structural $F_1$. Domain $F_1$ is unavailable for
untyped baselines (--). Best values per column in bold.}
\label{tab:e0_baselines}

\begin{tabular*}{\textwidth}{@{\extracolsep{\fill}}llccccccc@{}}
\toprule
\multirow{2}{*}{\textbf{Constructor}}
& \multirow{2}{*}{\textbf{Method}}
& \multirow{2}{*}{\textbf{nGED} $\downarrow$}
& \multirow{2}{*}{\textbf{Hall.} $\downarrow$}
& \multicolumn{5}{c}{\textbf{Structural} $F_1$ $\uparrow$} \\
\cmidrule(lr){5-9}
& & & &
\textbf{Tables} & \textbf{Attr.}
& \textbf{Domain} & \textbf{Entity}
& \textbf{Ref.} \\
\midrule

\multirow{4}{*}{\makecell[l]{CodeLlama\\13B}}
 & One-shot
 & 0.613 & 0.728 & 0.587 & 0.379 & -- & 0.130 & 0.034 \\
 & One-shot+CoT
 & 0.668 & 0.737 & 0.580 & 0.344 & -- & 0.047 & 0.004 \\
 & Few-shot
 & 0.540 & 0.589 & \textbf{0.600} & 0.461 & -- & 0.355 & 0.118 \\
 & Direct
 & 0.583 & 0.669 & 0.554 & 0.414 & 0.115 & 0.301 & 0.141 \\
\midrule

\multirow{4}{*}{\makecell[l]{Llama-3.1\\8B}}
 & One-shot
 & 0.602 & 0.714 & 0.564 & 0.399 & -- & 0.221 & 0.086 \\
 & One-shot+CoT
 & 0.670 & 0.764 & 0.553 & 0.334 & -- & 0.067 & 0.020 \\
 & Few-shot
 & 0.547 & 0.638 & 0.585 & 0.461 & -- & 0.373 & 0.146 \\
 & Direct
 & 0.615 & 0.696 & 0.572 & 0.377 & 0.121 & 0.199 & 0.104 \\
\midrule

\multirow{4}{*}{\makecell[l]{DeepSeek-Coder\\6.7B}}
 & One-shot
 & 0.718 & 0.722 & 0.406 & 0.321 & -- & 0.246 & 0.115 \\
 & One-shot+CoT
 & 0.848 & 0.838 & 0.251 & 0.169 & -- & 0.091 & 0.046 \\
 & Few-shot
 & \textbf{0.528} & \textbf{0.582}
 & \textbf{0.600} & \textbf{0.479} & --
 & \textbf{0.419} & \textbf{0.188} \\
 & Direct
 & 0.561 & 0.595 & 0.566 & 0.443
 & \textbf{0.150} & 0.391 & 0.184 \\

\bottomrule
\end{tabular*}
\end{table*}

Table~\ref{tab:e0_baselines} shows that few-shot prompting achieves lower nGED than Direct construction across all three open-weight models, and generally improves table, attribute, and entity recovery. One-shot CoT consistently reduces structural fidelity relative to one-shot prompting. These results indicate that prompting strategy substantially affects schema design performance, while referential integrity remains challenging across configurations.

\subsection{E1: Schema Verbalization into Natural Language}
\label{sec:e1}

\paragraph{Setup}
We evaluate five describer LLMs under basic (B) and detailed (D) prompting (\S\ref{sec:prompting_protocol}). The evaluation covers 145 \emph{Enterprise-145}, 381 DDL-matched \emph{RSchema}, and 239 \emph{SchemaBench} schemata. We measure description quality using corpus BLEU-4, ROUGE-L, and BERTScore-$F_1$ against gold natural-language descriptions, and schema coverage using SCR. Since \emph{Enterprise-145} lacks gold natural-language descriptions, only SCR is reported for this collection.

\begin{table*}[t]
\centering
\small
\setlength{\tabcolsep}{3pt}
\renewcommand{\arraystretch}{1.0}

\caption{E1: Schema verbalization under basic (B) and
detailed (D) prompting. Best values per dataset
and column are in bold.}
\label{tab:e1_s2t}

\begin{tabular}{@{}ll cc cc cc @{\hspace{18pt}} cc@{}}
\toprule
& &
\multicolumn{2}{c}{\textbf{BLEU-4} $\uparrow$} &
\multicolumn{2}{c}{\textbf{ROUGE-L} $\uparrow$} &
\multicolumn{2}{c}{\textbf{BERTScore} $\uparrow$} &
\multicolumn{2}{c}{\textbf{SCR} $\uparrow$} \\

\cmidrule(lr){3-4}
\cmidrule(lr){5-6}
\cmidrule(lr){7-8}
\cmidrule(lr){9-10}

\textbf{Dataset} &
\textbf{Describer} &
B & D & B & D & B & D & B & D \\

\midrule

\multirow{5}{*}{\emph{\textit{Enterprise-145}}}
& CodeLlama-13B
& -- & -- & -- & -- & -- & --
& 0.514 & 0.605 \\

& Llama-3.1-8B
& -- & -- & -- & -- & -- & --
& 0.447 & 0.636 \\

& DeepSeek-Coder
& -- & -- & -- & -- & -- & --
& \textbf{0.638} & 0.661 \\

& Claude Sonnet 4.6
& -- & -- & -- & -- & -- & --
& 0.472 & 0.882 \\

& GPT-5.4-mini
& -- & -- & -- & -- & -- & --
& 0.411 & \textbf{0.917} \\

\midrule

\multirow{5}{*}{\emph{\textit{RSchema}}}
& CodeLlama-13B
& \textbf{7.63} & 4.55
& 0.260 & 0.225
& \textbf{0.864} & 0.840
& 0.748 & 0.932 \\

& Llama-3.1-8B
& 6.37 & \textbf{6.76}
& \textbf{0.263} & \textbf{0.252}
& 0.862 & \textbf{0.863}
& 0.519 & 0.754 \\

& DeepSeek-Coder
& 2.90 & 2.56
& 0.226 & 0.214
& 0.832 & 0.829
& \textbf{0.963} & 0.973 \\

& Claude Sonnet 4.6
& 4.60 & 1.19
& 0.241 & 0.189
& 0.859 & 0.817
& 0.421 & 0.999 \\

& GPT-5.4-mini
& 3.18 & 1.52
& 0.244 & 0.185
& 0.860 & 0.816
& 0.365 & \textbf{1.000} \\

\midrule

\multirow{5}{*}{\emph{\textit{SchemaBench}}}
& CodeLlama-13B
& 3.79 & 2.34
& 0.168 & 0.145
& 0.819 & 0.802
& 0.556 & 0.758 \\

& Llama-3.1-8B
& \textbf{4.19} & \textbf{4.09}
& \textbf{0.174} & \textbf{0.175}
& \textbf{0.831} & \textbf{0.815}
& 0.404 & 0.605 \\

& DeepSeek-Coder
& 3.02 & 1.71
& 0.146 & 0.141
& 0.798 & 0.792
& \textbf{0.702} & 0.825 \\

& Claude Sonnet 4.6
& 2.09 & 1.07
& 0.167 & 0.128
& 0.817 & 0.785
& 0.386 & 0.944 \\

& GPT-5.4-mini
& 1.02 & 1.02
& 0.157 & 0.124
& 0.816 & 0.781
& 0.316 & \textbf{0.974} \\

\bottomrule
\end{tabular}
\end{table*}

\begin{figure*}[t]
\centering
\includegraphics[width=\linewidth]{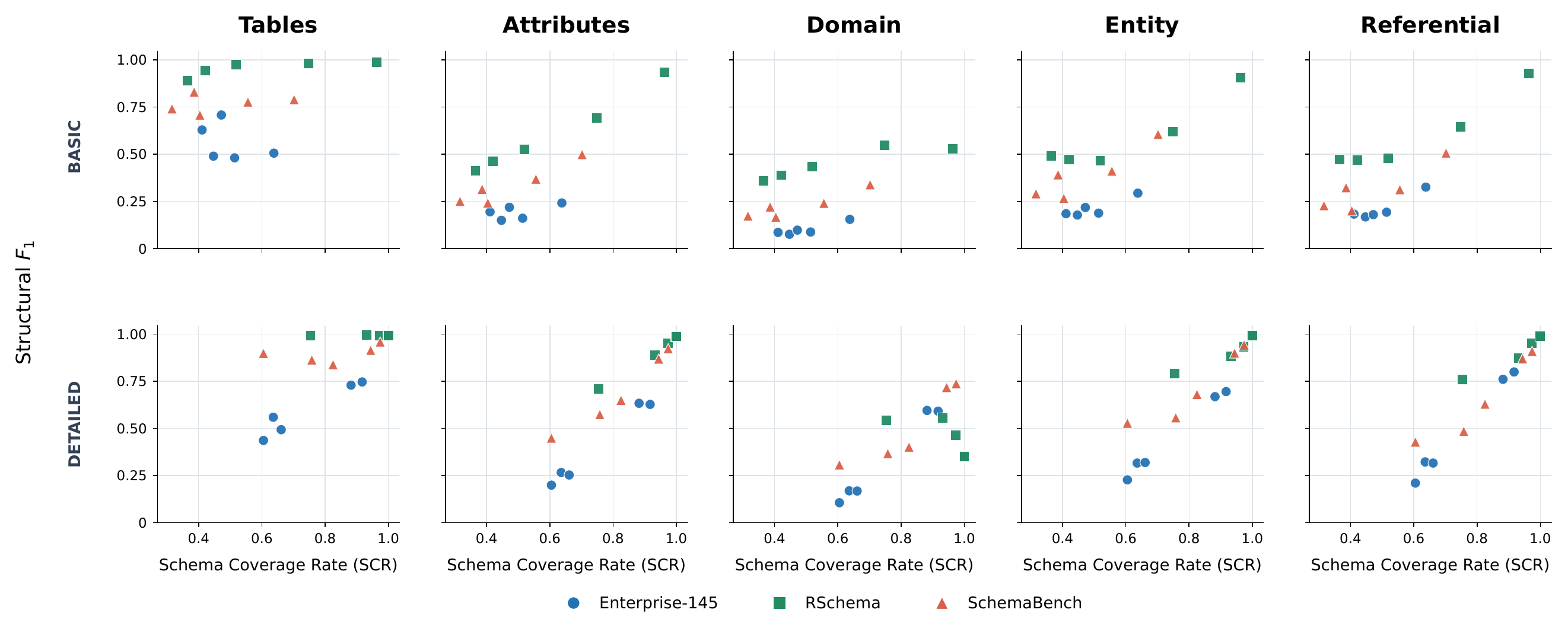}
\caption{E1--E2: Relationship between Schema Coverage Rate (SCR) and round-trip structural $F_1$ across five dimensions under basic (top) and detailed (bottom) prompting. Each panel contains 15 model-dataset observations from Enterprise-145, RSchema, and SchemaBench. High identifier coverage does not necessarily imply complete structural preservation, particularly for domain integrity.}
\label{fig:e2_scr}
\end{figure*}

\paragraph{Structural coverage vs. text similarity}
Table~\ref{tab:e1_s2t} shows that detailed prompting increases SCR for all five describers across the three datasets, while text similarity generally decreases. Claude Sonnet 4.6 and GPT-5.4-mini achieve near-complete identifier coverage on \emph{RSchema} and \emph{SchemaBench}. Gold descriptions tend to summarize requirements rather than list schema elements, which helps explain the lower similarity scores. SCR measures identifier mentions, not the correctness of types, keys, or relationships. Figure~\ref{fig:e2_scr} compares SCR with round-trip structural $F_1$ across these dimensions.

\subsection{E2: Round-Trip Preservation}
\label{sec:e2}

\paragraph{Setup}
We reconstruct the basic (B) and detailed (D) descriptions from E1 using a fixed GPT-4o-mini constructor (\S\ref{sec:prompting_protocol}). The evaluation includes 145 \emph{Enterprise-145}, 381 DDL-matched
\emph{RSchema}, and 239 \emph{SchemaBench} schemata. We report mean normalized graph edit distance (nGED) and micro-averaged $F_1$ for tables, attributes, domain, entity, and referential integrity.

\begin{figure*}[t]
\centering
\includegraphics[width=\linewidth]{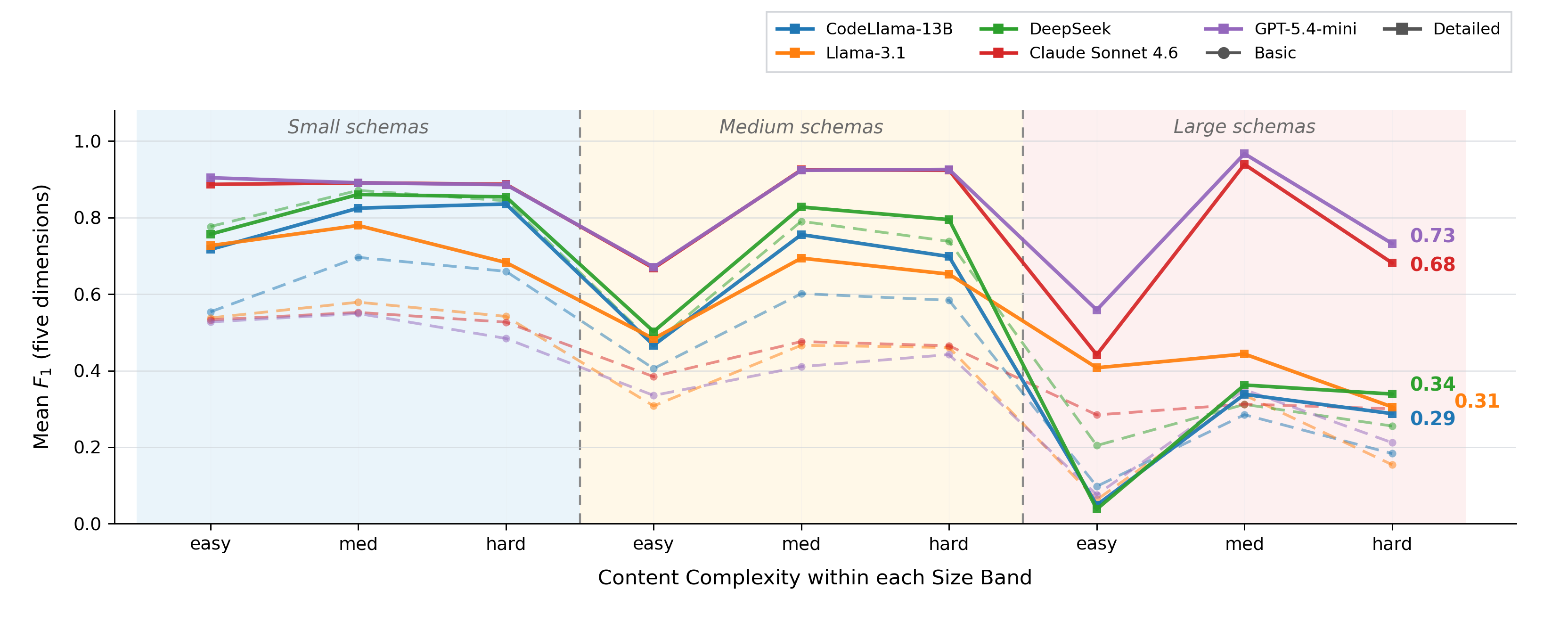}
\caption{E2: Mean structural $F_1$ across five dimensions by schema size and constraint complexity (765 schemata). Dashed and solid lines indicate basic and detailed prompting, respectively. Shaded regions denote size tiers. Constructor: GPT-4o-mini.}
\label{fig:e2_trajectory}
\end{figure*}

\begin{table*}[t]
\centering
\small
\setlength{\tabcolsep}{3.5pt}
\renewcommand{\arraystretch}{1.0}

\caption{E2: Round-trip preservation under basic (B) and detailed (D)
prompting with GPT-4o-mini as the constructor. Mean nGED ($\downarrow$)
and micro-averaged structural $F_1$ ($\uparrow$). Best values per
dataset and column are in bold.}
\label{tab:e2_roundtrip}

\resizebox{\textwidth}{!}{%
\begin{tabular}{@{}llcccccccccccc@{}}
\toprule
\multirow{3}{*}{\textbf{Dataset}}
& \multirow{3}{*}{\textbf{Describer}}
& \multicolumn{2}{c}{\textbf{nGED} $\downarrow$}
& \multicolumn{10}{c}{\textbf{Structural} $F_1$ $\uparrow$} \\
\cmidrule(lr){3-4}
\cmidrule(lr){5-14}

& & &
& \multicolumn{2}{c}{\textbf{Tables}}
& \multicolumn{2}{c}{\textbf{Attributes}}
& \multicolumn{2}{c}{\textbf{Domain}}
& \multicolumn{2}{c}{\textbf{Entity}}
& \multicolumn{2}{c}{\textbf{Referential}} \\

\cmidrule(lr){5-6}
\cmidrule(lr){7-8}
\cmidrule(lr){9-10}
\cmidrule(lr){11-12}
\cmidrule(lr){13-14}

& & B & D & B & D & B & D
& B & D & B & D & B & D \\
\midrule

\multirow{5}{*}{\emph{Enterprise-145}}
& CodeLlama-13B
& 0.714 & 0.625 & 0.481 & 0.436
& 0.161 & 0.199 & 0.088 & 0.106
& 0.188 & 0.227 & 0.193 & 0.210 \\

& Llama-3.1-8B
& 0.739 & 0.520 & 0.490 & 0.559
& 0.150 & 0.266 & 0.076 & 0.169
& 0.178 & 0.316 & 0.168 & 0.322 \\

& DeepSeek-Coder
& \textbf{0.561} & 0.505
& 0.506 & 0.493
& \textbf{0.242} & 0.253
& \textbf{0.155} & 0.168
& \textbf{0.294} & 0.319
& \textbf{0.326} & 0.316 \\

& Claude Sonnet 4.6
& 0.718 & \textbf{0.101}
& \textbf{0.708} & 0.729
& 0.219 & \textbf{0.633}
& 0.098 & \textbf{0.595}
& 0.218 & 0.668
& 0.180 & 0.760 \\

& GPT-5.4-mini
& 0.711 & 0.108
& 0.629 & \textbf{0.746}
& 0.195 & 0.627
& 0.086 & 0.591
& 0.185 & \textbf{0.695}
& 0.183 & \textbf{0.799} \\

\midrule

\multirow{5}{*}{\emph{RSchema}}
& CodeLlama-13B
& 0.364 & \textbf{0.339}
& 0.980 & \textbf{0.994}
& 0.692 & 0.889
& \textbf{0.547} & \textbf{0.555}
& 0.621 & 0.883
& 0.646 & 0.871 \\

& Llama-3.1-8B
& 0.461 & 0.348
& 0.975 & 0.991
& 0.526 & 0.710
& 0.435 & 0.543
& 0.467 & 0.790
& 0.479 & 0.758 \\

& DeepSeek-Coder
& \textbf{0.353} & 0.390
& \textbf{0.988} & 0.991
& \textbf{0.933} & 0.950
& 0.528 & 0.463
& \textbf{0.907} & 0.932
& \textbf{0.928} & 0.952 \\

& Claude Sonnet 4.6
& 0.501 & 0.468
& 0.945 & \textbf{0.994}
& 0.462 & 0.985
& 0.389 & 0.349
& 0.471 & \textbf{0.994}
& 0.469 & 0.988 \\

& GPT-5.4-mini
& 0.519 & 0.469
& 0.890 & 0.992
& 0.414 & \textbf{0.987}
& 0.360 & 0.351
& 0.491 & 0.991
& 0.473 & \textbf{0.989} \\

\midrule

\multirow{5}{*}{\emph{SchemaBench}}
& CodeLlama-13B
& 0.581 & 0.443
& 0.777 & 0.863
& 0.369 & 0.574
& 0.241 & 0.367
& 0.411 & 0.557
& 0.313 & 0.485 \\

& Llama-3.1-8B
& 0.647 & 0.497
& 0.708 & 0.898
& 0.242 & 0.449
& 0.167 & 0.307
& 0.267 & 0.528
& 0.201 & 0.428 \\

& DeepSeek-Coder
& \textbf{0.419} & 0.403
& 0.790 & 0.838
& \textbf{0.499} & 0.649
& \textbf{0.339} & 0.401
& \textbf{0.606} & 0.680
& \textbf{0.507} & 0.629 \\

& Claude Sonnet 4.6
& 0.622 & 0.212
& \textbf{0.830} & 0.914
& 0.315 & 0.868
& 0.221 & 0.717
& 0.392 & 0.899
& 0.323 & 0.869 \\

& GPT-5.4-mini
& 0.663 & \textbf{0.200}
& 0.741 & \textbf{0.958}
& 0.251 & \textbf{0.923}
& 0.173 & \textbf{0.737}
& 0.291 & \textbf{0.942}
& 0.228 & \textbf{0.909} \\

\bottomrule
\end{tabular}%
}
\end{table*}

\paragraph{Prompting and structural preservation}
Table~\ref{tab:e2_roundtrip} shows that detailed prompting generally improves round-trip fidelity, particularly for attributes, entity constraints, and foreign keys. The API-based describers achieve the
highest detailed prompt $F_1$ in several dimensions, but the smallest nGED on \emph{RSchema} is obtained by CodeLlama-13B. Domain integrity remains comparatively difficult to recover, and a more detailed description does not improve every model-dataset combination.

\paragraph{Schema complexity and failure modes}
Figure~\ref{fig:e2_trajectory} shows that structural fidelity generally declines with schema size, although the trend varies across describer LLM and constraint complexity. Detailed prompting improves preservation for most models, with Claude Sonnet 4.6 and GPT-5.4-mini maintaining comparatively high fidelity on larger schemata. These results indicate that detailed descriptions mitigate, but do not eliminate, complexity-dependent reconstruction errors.

\subsection{E3: Isolating the Constructor}
\label{sec:e3}

\paragraph{Setup}
Unlike E2, which fixes the constructor, E3 fixes GPT-5.4-mini detailed descriptions and varies the
constructor across five LLMs. We evaluate 765 DDL-backed schemata: 145 \emph{Enterprise-145}, 381 \emph{RSchema}, and 239 \emph{SchemaBench}. We report mean nGED, micro-averaged structural $F_1$, and four-dimensional hallucination over tables, attributes, entities, and referential integrity, excluding domain.

\begin{table*}[t]
\centering
\small
\setlength{\tabcolsep}{3pt}
\renewcommand{\arraystretch}{1.0}
\caption{E3: Constructor isolation using fixed GPT-5.4-mini
detailed descriptions ($N=765$). Mean nGED and
four-dimensional hallucination ($\downarrow$), and
micro-averaged structural $F_1$ ($\uparrow$).
Best values per dataset and metric in bold.}
\label{tab:e3_overall}

\begin{tabular*}{\textwidth}{@{\extracolsep{\fill}}llccccccc@{}}
\toprule
\multirow{2}{*}{\textbf{Dataset}}
& \multirow{2}{*}{\textbf{Constructor}}
& \multirow{2}{*}{\textbf{nGED} $\downarrow$}
& \multirow{2}{*}{\textbf{Hall.} $\downarrow$}
& \multicolumn{5}{c}{\textbf{Structural $F_1$ $\uparrow$}} \\
\cmidrule(lr){5-9}
& & & &
\textbf{Tables} & \textbf{Attr.}
& \textbf{Domain} & \textbf{Entity} & \textbf{Ref.} \\
\midrule

\multirow{5}{*}{\emph{Enterprise-145}}
& CodeLlama-13B
& 0.324 & 0.085 & 0.553 & 0.479
& 0.315 & 0.479 & 0.460 \\

& Llama-3.1-8B
& 0.340 & 0.071 & 0.596 & 0.294
& 0.277 & 0.489 & 0.528 \\

& DeepSeek-Coder
& 0.549 & 0.120 & 0.492 & 0.167
& 0.117 & 0.348 & 0.296 \\

& Claude Sonnet 4.6
& \textbf{0.047} & 0.027
& \textbf{0.758} & \textbf{0.703}
& \textbf{0.685} & \textbf{0.744}
& \textbf{0.847} \\

& GPT-5.4-mini
& 0.058 & \textbf{0.026}
& 0.753 & 0.683 & 0.664
& 0.720 & 0.816 \\

\midrule

\multirow{5}{*}{\emph{RSchema}}
& CodeLlama-13B
& \textbf{0.369} & 0.020
& 0.987 & 0.986 & 0.483
& 0.981 & 0.969 \\

& Llama-3.1-8B
& 0.493 & 0.020
& 0.989 & 0.986 & 0.343
& 0.976 & 0.970 \\

& DeepSeek-Coder
& 0.383 & 0.073
& 0.940 & 0.807 & \textbf{0.508}
& 0.806 & 0.802 \\

& Claude Sonnet 4.6
& 0.468 & \textbf{0.009}
& \textbf{0.994} & \textbf{0.991}
& 0.351 & \textbf{0.995}
& \textbf{0.994} \\

& GPT-5.4-mini
& 0.471 & 0.010
& 0.993 & 0.990 & 0.345
& \textbf{0.995} & 0.991 \\

\midrule

\multirow{5}{*}{\emph{SchemaBench}}
& CodeLlama-13B
& 0.272 & 0.054
& 0.760 & 0.697 & 0.527
& 0.714 & 0.623 \\

& Llama-3.1-8B
& 0.369 & 0.058
& 0.806 & 0.491 & 0.370
& 0.711 & 0.672 \\

& DeepSeek-Coder
& 0.550 & 0.096
& 0.649 & 0.251 & 0.184
& 0.523 & 0.423 \\

& Claude Sonnet 4.6
& \textbf{0.194} & \textbf{0.024}
& \textbf{0.964} & \textbf{0.945}
& \textbf{0.773} & \textbf{0.957}
& 0.930 \\

& GPT-5.4-mini
& 0.198 & 0.025
& 0.961 & 0.921 & 0.745
& 0.951 & \textbf{0.934} \\

\bottomrule
\end{tabular*}
\end{table*}

\paragraph{Constructor capability and domain errors}
Table~\ref{tab:e3_overall} shows that constructor choice substantially affects structural fidelity even with identical descriptions. Claude Sonnet 4.6 achieves the lowest nGED on \emph{Enterprise-145} and \emph{SchemaBench}, whereas CodeLlama-13B achieves the lowest nGED on \emph{RSchema}. On \emph{RSchema}, domain $F_1$ remains substantially lower than other structural dimensions, indicating weaker joint data-type and nullability fidelity, potentially influenced by dataset-specific annotation conventions.

\subsection{E4: Masking and Inferability}
\label{sec:e4}

\paragraph{Setup}
E4 evaluates recovery of hidden schema information under five masks: $M_T$ (tables), $M_A$ (non-key attributes), $M_D$ (domain properties), $M_E$ (entity constraints), and $M_R$ (foreign keys). GPT-4o-mini reconstructs masked DDL directly or after verbalization by an LLM describer. Recall is micro-averaged over available scored outputs, grouped by heuristic inferability.

For $M_D$, the historical masking and inferability protocol is retained, but scoring uses joint data-type family and nullability matching. Domain evaluation is restricted to genuine DDL from \emph{Enterprise-145} and \emph{SchemaBench}; synthetic \emph{RSchema} domain references are excluded. The historical heuristic assigns no inferable domain labels to \emph{Enterprise-145}. We also exclude 66 cases without Domain Direct outputs (1,663 eligible gold elements).

\begin{table}[t]
\centering
\small
\setlength{\tabcolsep}{4pt}
\renewcommand{\arraystretch}{1.0}

\caption{E4: Direct recall by heuristic inferability (GPT-4o-mini; observed-case outputs). Parentheses show eligible gold element counts.}
\label{tab:e4_split}

\begin{tabular}{@{}lcc@{}}
\toprule
\textbf{Masked dimension}
& \textbf{Inferable}
& \textbf{Non-inferable} \\
\midrule

$M_T$ Tables
& 0.350 (2{,}555)
& 0.197 (4{,}272) \\

$M_A$ Attributes
& 0.611 (9{,}254)
& 0.187 (42{,}594) \\

$M_D$ Domain
& 0.409 (1{,}574)
& 0.479 (35{,}573) \\

$M_E$ Entity
& 0.872 (2{,}161)
& 0.635 (2{,}608) \\

$M_R$ Referential
& 0.687 (4{,}214)
& 0.288 (1{,}246) \\

\bottomrule
\end{tabular}
\end{table}

\paragraph{Evidence-guided reconstruction}
Table~\ref{tab:e4_split} shows that inferable elements are recovered more frequently than non-inferable ones for tables, attributes, entity constraints, and foreign keys. Recovery is strongest for inferable entity constraints, whereas anonymized table names remain difficult even when naming and relational cues are present. The substantial gaps for attributes and foreign keys suggest that visible structural evidence helps reconstruct missing schema information. However, non-inferable recovery remains incomplete, and the heuristic classification does not establish whether successful predictions arise from logical reasoning or learned design conventions.

\begin{table}[t]
\centering
\small
\setlength{\tabcolsep}{4pt}
\renewcommand{\arraystretch}{1.05}

\caption{E4: Paired Direct and Round-trip recall of
heuristic-inferable elements. Round-trip recall
is averaged over available describers.
Retention is Round-trip/Direct recall.
Domain uses joint data-type and nullability matching.}
\label{tab:e4_recovery}

\begin{tabular}{@{}lcccc@{}}
\toprule
\textbf{Dimension}
& \textbf{Gold $N$}
& \textbf{Direct}
& \textbf{RT}
& \textbf{Retention} \\
\midrule

$M_T$ Tables
& 2{,}555 & 0.350 & 0.120 & 0.344 \\

$M_A$ Attributes
& 9{,}254 & 0.611 & 0.487 & 0.797 \\

$M_D$ Domain
& 1{,}574 & 0.409 & 0.032 & 0.078 \\

$M_E$ Entity
& 2{,}087 & 0.878 & 0.625 & 0.712 \\

$M_R$ Referential
& 4{,}197 & 0.688 & 0.444 & 0.645 \\

\bottomrule
\end{tabular}
\end{table}

\paragraph{Structural evidence and verbalization loss}
Table~\ref{tab:e4_recovery} shows that verbalization reduces recovery across all five dimensions. Attributes retain the highest proportion of Direct recall (0.797), followed by entity constraints (0.712) and foreign keys (0.645). Domain exhibits the lowest retention (0.078), with Round-trip recall falling from 0.409 to 0.032. This substantial loss in joint data-type and nullability recovery indicates
that verbalization can discard schema-critical information even when structural evidence is available.


\subsection{E5: Scalability and Cost}
\label{sec:e5}

\paragraph{Setup}
We evaluate round-trip inference efficiency using GPT-5.4-mini and Claude Sonnet 4.6 as describers, with GPT-4o-mini as the fixed constructor across 30 schemas (10 per dataset). We measure end-to-end latency, token consumption, and API cost per schema under basic and detailed prompting, including both verbalization and reconstruction stages.

\begin{table}[t]
\centering
\caption{E5: Mean round-trip latency, token usage, and API cost under basic (B) and detailed (D) prompting.}
\label{tab:e5_cost}

\footnotesize
\setlength{\tabcolsep}{2.2pt}
\renewcommand{\arraystretch}{1.05}

\resizebox{\columnwidth}{!}{%
\begin{tabular}{@{}llcccccc@{}}
\toprule
\multirow{2}{*}{\textbf{Dataset}}
& \multirow{2}{*}{\textbf{Describer}}
& \multicolumn{2}{c}{\textbf{Latency (s)} $\downarrow$}
& \multicolumn{2}{c}{\textbf{Tokens} $\downarrow$}
& \multicolumn{2}{c}{\textbf{Cost (USD)} $\downarrow$} \\
\cmidrule(lr){3-4}
\cmidrule(lr){5-6}
\cmidrule(lr){7-8}
& & B & D & B & D & B & D \\
\midrule

\multirow{2}{*}{\emph{\textit{RSchema}}}
& GPT-5.4-mini
& 68.4 & 65.8 & 5750 & 7444 & 0.0080 & 0.0116 \\
& Claude Sonnet 4.6
& 72.5 & 86.3 & 6293 & 8410 & 0.0186 & 0.0343 \\

\midrule
\multirow{2}{*}{\emph{\textit{SchemaBench}}}
& GPT-5.4-mini
& 108.8 & 130.1 & 9496 & 12451 & 0.0181 & 0.0249 \\
& Claude Sonnet 4.6
& 134.8 & 145.0 & 10243 & 13022 & 0.0341 & 0.0554 \\

\midrule
\multirow{2}{*}{\emph{\textit{Enterprise-145}}}
& GPT-5.4-mini
& 132.5 & 139.4 & 11231 & 13700 & 0.0194 & 0.0250 \\
& Claude Sonnet 4.6
& 165.9 & 164.9 & 12056 & 14248 & 0.0357 & 0.0524 \\

\bottomrule
\end{tabular}%
}
\end{table}

\paragraph{Inference efficiency}
Table~\ref{tab:e5_cost} shows that detailed prompting increases token consumption and API cost across all configurations, while latency varies with the model and dataset. Larger schemata incur higher inference costs due to longer descriptions and reconstruction outputs. Claude Sonnet 4.6 is consistently more expensive than GPT-5.4-mini, reflecting differences in token usage and API pricing. These results highlight the computational cost of preserving structural detail during round-trip reconstruction.

\paragraph{Evaluation limitations}
Our results measure reference-based structural fidelity rather than functional schema equivalence. The experiments use model generations, and we do not estimate run-to-run variability. Small differences between models should therefore be interpreted cautiously. The controlled round-trip comparisons are
conditional on fixed models and do not estimate describer-constructer interactions.

\section{Recommendations}
\label{sec:recommendations}

We provide several practical recommendations for generating relational schemas with LLMs and describing them.

\spara{Specify structural requirements explicitly.}
When describing a schema, prompts should explicitly request tables, attributes, data types, keys, and foreign-key relationships. Detailed prompting substantially improves structural preservation, whereas brief descriptions often omit schema-critical information. Likewise, when generating a schema from a natural language description, missing structural details should be stated explicitly.

\spara{Validate domain and referential constraints.}
Even when LLMs recover tables and relationships accurately, they often struggle with data types and nullability. Domain constraints, primary keys, and foreign keys should therefore be validated before deployment. The masking experiment shows that LLMs often infer plausible relationships from column names and existing constraints, but these need not reflect the intended design and should be checked against application requirements.

\spara{Evaluate structural fidelity beyond text similarity.}
Machine translation metrics such as BLEU, ROUGE-L, and BERTScore do not reliably indicate whether a description preserves the original schema. We recommend evaluating schema comprehension through structural reconstruction, with fidelity reported separately for tables, attributes, domain, entity, and referential integrity. Schema Coverage Rate (SCR) additionally measures identifier preservation.

\spara{Account for schema complexity and model roles.}
Structural preservation deteriorates as schema size grows, even under detailed prompting. A model that describes a schema well is not necessarily a good generator, so model selection should treat the two tasks independently rather than rely on general LLM rankings. For large or complex schemata, additional structural validation is needed to identify omitted attributes and relationships.

\section{Conclusion and Future Work}
\label{sec:conclusion}

We presented \textsf{SchemaBridge}, a benchmark for evaluating relational schema design, comprehension, and structural preservation across 902 schemata. Our experiments show that given natural language requirements often underspecify schema-critical information, while detailed descriptions generally improve round-trip fidelity. We further demonstrate that text-similarity metrics do not reliably reflect structural correctness and that models recover relational constraints more effectively than domain information. These results encourage the importance of explicit constraint-aware prompting, per-dimension evaluation, and validation of inferred schema elements.

Future work will explore agentic AI frameworks, external schema-aware knowledge bases, and human-in-the-loop evaluation with LLM-as-a-Judge to assess semantic and structural fidelity. We also aim to develop an end-to-end framework integrating schema generation, verbalization, reconstruction, and validation, with unified metrics for evaluating both directions of the complete round-trip process.

\section*{AI-Generated Content Acknowledgment}

OpenAI ChatGPT and Cursor were used selectively to assist with experimental code development, debugging, writing the paper, and the preparation of certain figures. All research decisions, experimental results, and interpretations were verified by the authors.



\clearpage
\bibliographystyle{IEEEtran}
\bibliography{references}

\end{document}